\documentclass[10pt,a4paper,twoside]{article}

\usepackage{vmargin,fancyheadings,multicol,multirow,ifthen,cite,
            graphicx,wrapfig,calc,dcolumn,apalike,setspace,
            boxedminipage,rotating,textcomp,picinpar,longtable,
            url,amsmath,amssymb,ogonek} 

\usepackage[tight]{subfigure}
\usepackage{imc2024}
\renewcommand{\PageTitle}{Proceedings of the IMC, Kutn\'a Hora, 2024. Published in WGN.}

\newcommand {\0}  {\phantom{0}}

\begin{document}
\SetPaperBodyFont

\begin{IMCpaper}{
\title{Meteor observations as a tool to constrain cosmogonic models of the Solar System}
\author{Vitalii Kuksenko$^1$ and Juraj T\'oth$^1$
        \thanks{$^1\,$Department of Astronomy, Physics of the Earth, and Meteorology, \\
                      Faculty of mathematics, physics and informatics, Comenius University, Bratislava, Slovakia \\
                     \texttt{vitalii.kuksenko@fmph.uniba.sk}\\}}%
\abstract{Recent observations of small bodies of the Solar System showed
  evidence of the presence of refractory (asteroidal) material in the Oort
  cloud. Different models of the origin of the Solar System predict different
  numbers of rocky objects in the Oort cloud, meaning that measurement of
  this population can be used as an observational constraint for cosmogonic
  models. The aim of our work is to study how the data obtained from meteor
  observations can be used as a tool for distinguishing among the existing
  cosmogonic models. We investigated two meteor databases collected by the
  cameras of the All-Sky Meteor Orbit System (AMOS) located in the Canary
  Islands and in Chile. We describe methodology and results of the search for
  unusually strong rocky meteoroids on cometary orbits with the origin in the
  Oort cloud. These data will be used to calculate the fluxes of meteors of
  different compositions in order to constrain the ratio of icy and rocky
  components of the Oort cloud.  For the flux determination, we estimate the
  observational time and effective area of the AMOS system.}}%
\index{Kuksenko V.}%
\index{T\'oth J.}%
\vspace*{-3\baselineskip}

\section{Introduction}

It is now believed that the basic physics of planetary formation is generally
understood. However, during the last decades, evidence has accumulated that a
classical cosmogonic model for the Solar System cannot explain all its
constraints in detail \cite{raymond09}. Several improvements were proposed to
solve existing problems, the most successful were based on two revolutionary
concepts: planetary migrations \cite{tsiganis05,walsh11} and pebble accretion
\cite{lj12}. Since there are many new cosmogonic models, we need some tool to
test their validity and accuracy. One observational constraint suitable for
this purpose is the ice-to-rock ratio of the Oort cloud \cite{shannon15}.
Observations of small bodies of the Solar System revealed peculiar objects
known as Manx comets \cite{meech16} that represent macroscopic refractory
material originating from the Oort cloud. Recent detections of large rocky
meteoroids coming from cometary orbits \cite{sb99,vida23} show that meteor
observations can be used to indirectly measure the population of asteroidal
bodies and thus the ice-to-rock ratio of the Oort cloud \cite{vida23}.

The goal of our work is to use our meteor data to calculate fluxes of fragile
cometary and strong refractory meteoroids originating from the Oort cloud as
a proxy for the ice-to-rock ratio. The results of real observations can then
be compared with theoretical predictions of different cosmogonic models of
the Solar System \cite{meech16,shannon15,shannon19,vida23} to assess which of
them correctly reproduce the population of Oort cloud objects.

\section{Methodology}

\subsection*{Data}

\begin{figure*}[!t]
\centering
\includegraphics[width = \columnwidth]{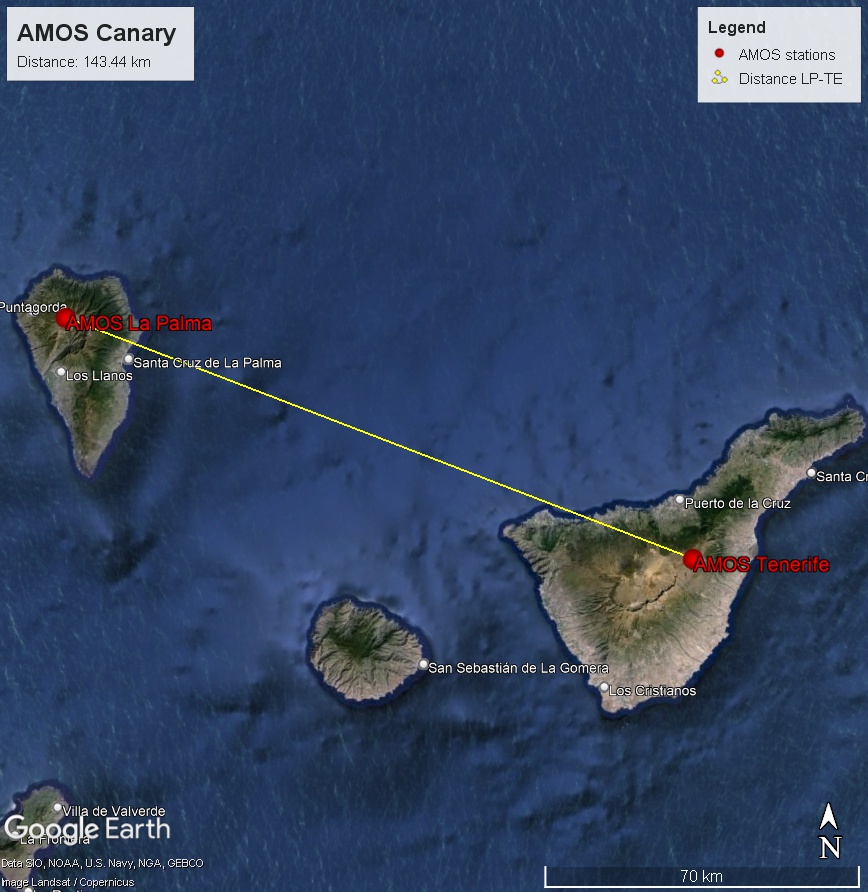}%
\includegraphics[width = \columnwidth]{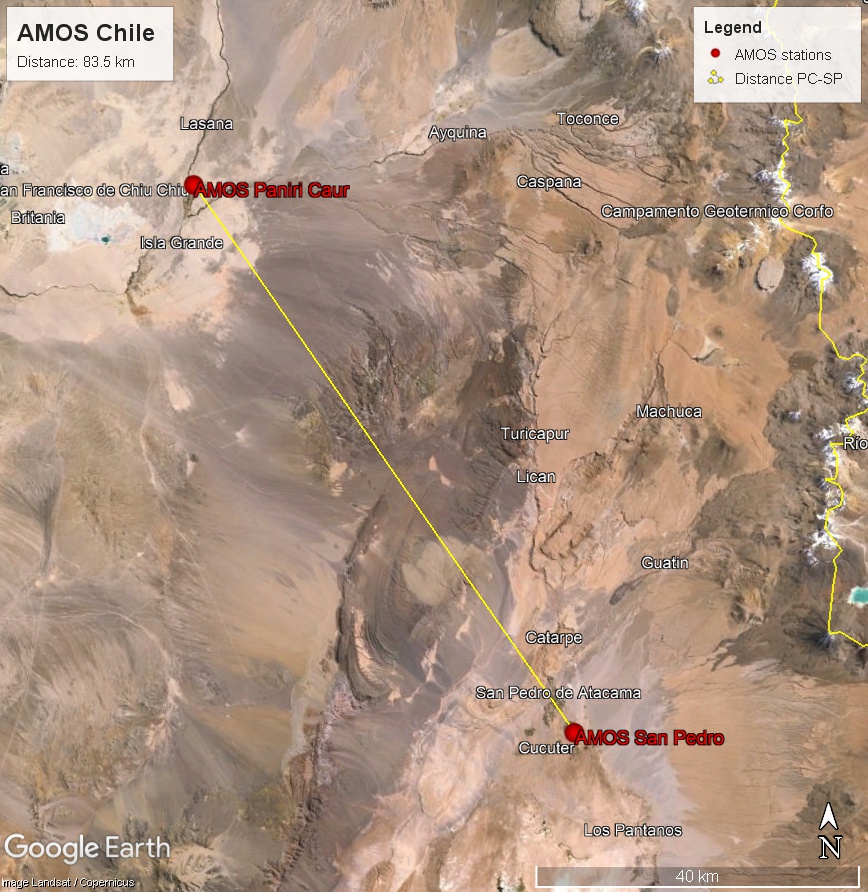}%
\caption{Geometry of AMOS Canary (left) and AMOS Chile (right) stations used
  in this work.}
\label{fig:map}
\end{figure*}

\begin{table*}[!t]
\caption{Geographical coordinates, altitudes and distances between the AMOS
  stations used in this work.}%
\label{tab:coord}
\vspace*{6pt}
\centering
\begin{tabular}{lcccc}
\hline
\0\0\0\0\0Station   & Longitude  & Latitude    & Altitude & Distance \\
                    &            &             &          & between \\
                    &            &             &          & stations \\
\hline
La Palma (Canary)   & $-$17\dg88 & \0\,28\dg76 &   2339 m & \multicolumn{1}{|c}{143.44 km} \\
Tenerife (Canary)   & $-$16\dg51 & \0\,28\dg30 &   2416 m & \multicolumn{1}{|c}{} \\
\hline
Paniri Caur (Chile) & $-$68\dg64 &  $-$22\dg34 &   2535 m & \multicolumn{1}{|c} {83.5  km} \\
San Pedro (Chile)   & $-$68\dg18 &  $-$22\dg95 &   2404 m & \multicolumn{1}{|c}{} \\
\hline
\end{tabular}
\end{table*}

In our work, we analyzed two databases collected by the All-Sky Meteor Orbit
System (AMOS) cameras \cite{toth11,toth15}. We used data from cameras located
in the Canary Islands and Chile (Figure~\ref{fig:map};
Table~\ref{tab:coord}).

\subsection*{Selection criteria}

After cleaning the databases from non-physical cases, we used a
multiparameter approach to search for dense meteoroids on cometary (HTC/LPC:
Halley-type comets or long period comets) orbits with $T_J < 2$, where
\begin{equation}
T_J = \frac{a_J}{a} + 2\sqrt{\frac{a}{a_J}(1-e^2)}\cos i
\end{equation}
is\, the\, Tisserand\, parameter\, with\, respect\, to\, Jupiter
\cite{kresak69,levison96}, \, $a$, $e$, $i$\, are\, orbital
elements of a meteoroid and $a_J$ is semi-major axis of Jupiter.  Material
properties were assessed by calculating empirical parameters, which are
defined as
\begin{equation}
K_B = \log\rho_\mathrm{b} + 2.5 \log v_\infty - 0.5 \log \cos z_R
\end{equation}
\begin{equation}
P_E = \log\rho_\mathrm{e}-0.42\log m_\infty+1.49\log v_\infty -1.29\log\cos z_R
\end{equation}
\begin{equation}
P_{E,\mathrm{mod}} = P_E - \log v_\infty + 1.5
\end{equation}
where $\rho$ is the atmospheric density at the beginning/end height of a
meteor, $v_\infty$ is the beginning velocity, $z_R$ is zenith distance of the
apparent radiant of the meteor and $m_\infty$ is the photometric mass
\cite{ceplecha58,cm76,borovicka22II}. Atmospheric densities were taken from
the atmospheric model NRLMSISE-00 \cite{picone02}. Note that we need to use
proper units as defined in the original papers.

We looked for objects of strong material types based on these parameters
(taking the value $P_{E,\mathrm{mod}} > -4.6$ as a boundary between cometary
and asteroidal material).  Additionally, we checked meteoroids for their
beginning velocity (the faster, the denser) and penetration heights (the
lower, the denser). Since we were interested in larger objects, we set the
lower mass limit for meteoroids to 1 g. The results of the search are
provided in Tables \ref{tab:ast} and \ref{tab:com}.

\begin{table*}[!t]
\caption{Properties of selected asteroidal (rocky) meteors on cometary orbits
  from two databases. Columns from left to right: name of the meteor and
  database, end height, beginning velocity, semi-major axis, eccentricity,
  inclination, longitude of ascending node, argument of perihelion, Tisserand
  parameter, $K_B$ parameter, $P_E$ parameter, modified $P_E$ parameter,
  photometric mass (calculated according to Pecina and Ceplecha, 1983)
  \nocite{pc83}. Below the values are corresponding uncertainties.}%
\label{tab:ast}
\vspace*{6pt}
\centering
\begin{tabular}{@{}c@{\0}c@{\0}c@{\0}c@{\0}c@{\0}c@{\0}c@{\0}c@{\0}c@{\0}c@{\0}c@{\0}c@{\0}c@{}}
\hline
Meteor name & $h_\mathrm{end}$ & $v_\mathrm{beg}$ & $a$  & $e$       & $i$       & $\Omega$  & $\omega$  & $T_J$  & $K_B$ & $P_E$ & $P_{E,\mathrm{mod}}$ & $m_\mathrm{phot}$ \\
and database      &  [km]     & [km/s]    & [au]         &           & [ \g\ ]   & [ \g\ ]   & [ \g\ ]   &           &           &           &           & [g] \\
\hline
M20150916\_023732 & 70.12     & 67.82     & 4.47         & 0.83      & 166.74    & 352.72    &  62.99    &    0.16   & 8.05      & $-$4.15   & $-$4.48   & 1.02 \\
(Canary)          & $\pm$0.3  & $\pm$0.34 & $\pm$0.53    & $\pm$0.02 & $\pm$0.34 & $\pm$0.00 & $\pm$1.51 & $\pm$0.12 & $\pm$0.01 & $\pm$0.01 & $\pm$0.01 & $\pm$0.08 \\
M20151107\_001720 & 58.57     & 37.25     & 3.95         & 0.94      & 17.38     & 44.05     & 123.62    &    1.90   & 8.17      & $-$4.15   & $-$4.22   & 2.56 \\
(Canary)          & $\pm$1.52 & $\pm$1.29 & $\pm$4.39    & $\pm$0.02 & $\pm$2.13 & $\pm$0.00 & $\pm$0.98 & $\pm$1.15 & $\pm$0.04 & $\pm$0.04 & $\pm$0.04 & $\pm$0.5 \\
M20160516\_043403 & 62.03     & 45.55     & 31.46        & 0.99      & 16.32     & 55.53     & 326.53    &    0.51   & 8.22      & $-$4.41   & $-$4.57   & 1.33 \\
(Chile)           & $\pm$0.25 & $\pm$1.84 & $\pm$27.29   & $\pm$0.01 & $\pm$7.64 & $\pm$0.00 & $\pm$1.59 & $\pm$1.01 & $\pm$0.04 & $\pm$0.04 & $\pm$0.03 & $\pm$0.19 \\
M20160815\_000132 & 36.51     & 22.5      & 5.9          & 0.85      & 20.49     & 142.38    & 221.38    &    1.95   & 8.97      & $-$3.73   & $-$3.58   & 18.28 \\
(Chile)           & $\pm$0.69 & $\pm$0.31 & $\pm$1.58    & $\pm$0.02 & $\pm$0.51 & $\pm$0.00 & $\pm$0.96 & $\pm$0.23 & $\pm$0.02 & $\pm$0.17 & $\pm$0.17 & $\pm$10.46 \\
M20160927\_042913 & 51.62     & 21.73     & 20.69        & 0.96      & 10.82     & 184.26    & 226.66    &    1.36   & 8.27      & $-$4.54   & $-$4.38   & 23.87 \\
(Chile)           & $\pm$0.58 & $\pm$0.4  & $\pm$121.14  & $\pm$0.03 & $\pm$0.73 & $\pm$0.00 & $\pm$0.66 & $\pm$1.85 & $\pm$0.02 & $\pm$3.95 & $\pm$3.95 & $\pm$481.84 \\
M20161008\_075401 & 63.25     & 26.99     & 6.31         & 0.85      & 36.11     & 15.24     &  25.32    &    1.77   & 7.53      & $-$4.62   & $-$4.55   & 2.6 \\
(Chile)           & $\pm$0.21 & $\pm$0.15 & $\pm$0.43    & $\pm$0.01 & $\pm$0.15 & $\pm$0.00 & $\pm$0.07 & $\pm$0.11 & $\pm$0.01 & $\pm$0.02 & $\pm$0.02 & $\pm$0.24 \\
M20161103\_063642 & 65.75     & 35.99     & 9.48         & 0.9       & 52.66     & 41.05     & 327.99    &    1.26   & 7.85      & $-$4.42   & $-$4.48   & 5.94 \\
(Chile)           & $\pm$0.32 & $\pm$0.91 & $\pm$544.42  & $\pm$0.04 & $\pm$1.21 & $\pm$0.00 & $\pm$0.51 & $\pm$1.56 & $\pm$0.03 & $\pm$0.03 & $\pm$0.03 & $\pm$0.95 \\
M20161113\_063540 & 57.02     & 43.39     & 4.37         & 0.97      & 37.18     & 51.09     & 141.32    &    1.54   & 8.56      & $-$3.87   & $-$4.01   & 2.15 \\
(Chile)           & $\pm$0.17 & $\pm$0.54 & $\pm$0.43    & $\pm$0.01 & $\pm$1.21 & $\pm$0.00 & $\pm$1.03 & $\pm$0.33 & $\pm$0.01 & $\pm$0.02 & $\pm$0.02 & $\pm$0.2 \\
M20161120\_083311 & 67.42     & 32.06     & 7.09         & 0.93      & 26.54     & 58.23     &  88.49    &    1.53   & 7.2       & $-$4.48   & $-$4.49   & 2.72 \\
(Chile)           & $\pm$0.36 & $\pm$0.36 & $\pm$1.54    & $\pm$0.01 & $\pm$0.51 & $\pm$0.00 & $\pm$0.49 & $\pm$0.22 & $\pm$0.01 & $\pm$0.02 & $\pm$0.02 & $\pm$0.29 \\
\hline
\end{tabular}
\end{table*}

\begin{table*}[!t]
\caption{Properties of selected cometary meteors on cometary orbits from two
  databases. Columns are the same as in Table~\protect\ref{tab:ast}. Below
  the values are corresponding uncertainties.}%
\label{tab:com}
\vspace*{6pt}
\centering
\begin{tabular}{@{}c@{\0}c@{\0}c@{\0}c@{\0}c@{\0}c@{\0}c@{\0}c@{\0}c@{\0}c@{\0}c@{\0}c@{\0}c@{}}
\hline
Meteor name & $h_\mathrm{end}$ & $v_\mathrm{beg}$ & $a$  & $e$       & $i$       & $\Omega$  & $\omega$  & $T_J$  & $K_B$ & $P_E$ & $P_{E,\mathrm{mod}}$ & $m_\mathrm{phot}$ \\
and database      &  [km]     & [km/s]    & [au]         &           & [ \g\ ]   & [ \g\ ]   & [ \g\ ]   &           &           &           &           & [g] \\
\hline
M20150611\_000233 & 91.53     & 52.3      & 5.03         & 0.83      &  93.71    &  79.57    & 225.65    &    0.99   & 6.58      & $-$6.03   & $-$6.25   & 4.62 \\
(Canary)          & $\pm$0.74 & $\pm$0.38 & $\pm$0.86    & $\pm$0.03 & $\pm$0.29 & $\pm$0.00 & $\pm$1.17 & $\pm$0.32 & $\pm$0.01 & $\pm$0.02 & $\pm$0.02 & $\pm$0.52 \\
M20150617\_013035 & 90.16     & 51.24     & 59.61        & 0.98      &  87.34    &  85.37    & 177.61    &    0.17   & 6.73      & $-$5.82   & $-$6.03   & 1.22 \\
(Canary)          & $\pm$0.39 & $\pm$0.36 & $\pm$6538.76 & $\pm$0.02 & $\pm$0.64 & $\pm$0.00 & $\pm$0.75 & $\pm$2.84 & $\pm$0.01 & $\pm$0.02 & $\pm$0.02 & $\pm$0.13 \\
M20150623\_223239 & 92.02     & 51.62     & --           & 1.02      &  86.98    &  91.93    & 222.38    &      --   & 6.74      & $-$5.77   & $-$5.98   & 1.72 \\
(Canary)          & $\pm$0.21 & $\pm$0.3  &              & $\pm$0.02 & $\pm$0.21 & $\pm$0.00 & $\pm$0.41 &           & $\pm$0.01 & $\pm$0.02 & $\pm$0.02 & $\pm$0.19 \\
M20150716\_015659 & 79.57     & 65.08     & 2.35         & 0.71      & 177.46    & 293.13    &  79.29    &    1.27   & 7.27      & $-$4.85   & $-$5.16   & 3.61 \\
(Canary)          & $\pm$0.51 & $\pm$0.69 & $\pm$0.26    & $\pm$0.03 & $\pm$0.27 & $\pm$0.01 & $\pm$3.83 & $\pm$0.2  & $\pm$0.01 & $\pm$0.02 & $\pm$0.01 & $\pm$0.28 \\
M20150720\_032355 & 73.64     & 61.45     & --           & 1.07      & 121.76    & 296.92    &  90.7     &      --   & 6.63      & $-$4.8    & $-$5.09   & 21.67 \\
(Canary)          & $\pm$0.38 & $\pm$0.57 &              & $\pm$0.03 & $\pm$0.34 & $\pm$0.00 & $\pm$1.3  &           & $\pm$0.01 & $\pm$0.01 & $\pm$0.01 & $\pm$1.39 \\
M20150724\_012753 & 91.27     & 64.89     & 8.72         & 0.95      & 165.97    & 120.65    & 278.95    & $-$0.19   & 6.88      & $-$5.71   & $-$6.03   & 1.41 \\
(Canary)          & $\pm$0.32 & $\pm$0.23 & $\pm$1.67    & $\pm$0.01 & $\pm$0.15 & $\pm$0.00 & $\pm$1.02 & $\pm$0.09 & $\pm$0.00 & $\pm$0.02 & $\pm$0.02 & $\pm$0.17 \\
M20150724\_045540 & 77.1      & 52.44     & 23.31        & 0.96      &  91.26    & 120.8     & 162.81    &    0.22   & 7.25      & $-$4.86   & $-$5.08   & 1.04 \\
(Canary)          & $\pm$0.45 & $\pm$0.36 & $\pm$161.2   & $\pm$0.04 & $\pm$0.31 & $\pm$0.00 & $\pm$2.61 & $\pm$0.38 & $\pm$0.01 & $\pm$0.02 & $\pm$0.02 & $\pm$0.09 \\
M20150728\_024025 & 86.25     & 56.33     & 706.95       & 0.99      & 100.07    & 124.53    & 164.53    & $-$0.18   & 7.12      & $-$6.97   & $-$7.22   & 5075.71 \\
(Canary)          & $\pm$0.42 & $\pm$1.25 & $\pm$973.86  & $\pm$0.09 & $\pm$1.33 & $\pm$0.00 & $\pm$4.58 & $\pm$7.09 & $\pm$0.03 & $\pm$0.03 & $\pm$0.03 & $\pm$679.53 \\
M20150728\_050622 & 87.41     & 66.75     & --           & 1.06      & 141.31    & 124.63    & 113.86    &      --   & 7.75      & $-$5.39   & $-$5.71   & 1.25 \\
(Canary)          & $\pm$0.39 & $\pm$1.78 &              & $\pm$0.14 & $\pm$0.65 & $\pm$0.00 & $\pm$3.4  &           & $\pm$0.03 & $\pm$0.04 & $\pm$0.04 & $\pm$0.26 \\
M20150728\_051117 & 82.21     & 55.24     & 15.13        & 0.93      &  99.04    & 124.63    & 170.83    &    0.18   & 7.01      & $-$5.48   & $-$5.72   & 5.76 \\
(Canary)          & $\pm$0.25 & $\pm$0.33 & $\pm$121.82  & $\pm$0.03 & $\pm$0.14 & $\pm$0.00 & $\pm$0.31 & $\pm$0.38 & $\pm$0.01 & $\pm$0.02 & $\pm$0.02 & $\pm$0.55 \\
M20150801\_051357 & 73.9      & 54.24     & --           & 1.02      &  63.08    & 308.46    & 152.96    &      --   & 7.19      & $-$5.63   & $-$5.87   & 576.57 \\
(Canary)          & $\pm$0.54 & $\pm$2.27 &              & $\pm$0.01 & $\pm$8.69 & $\pm$0.00 & $\pm$0.64 &           & $\pm$0.05 & $\pm$0.1  & $\pm$0.09 & $\pm$291.63 \\
M20150802\_015447 & 89.1      & 60.34     & 29.35        & 0.97      & 112.41    & 129.28    & 156.77    & $-$0.26   & 7.27      & $-$6.37   & $-$6.65   & 121.34 \\
(Canary)          & $\pm$0.19 & $\pm$0.35 & $\pm$243.85  & $\pm$0.03 & $\pm$0.26 & $\pm$0.00 & $\pm$0.61 & $\pm$1.41 & $\pm$0.01 & $\pm$0.11 & $\pm$0.11 & $\pm$71.05 \\
M20150803\_041252 & 97.63     & 57.4      & 5.77         & 0.83      & 107.63    & 130.33    & 161.18    &    0.57   & 6.52      & $-$6.35   & $-$6.61   & 1.03 \\
(Canary)          & $\pm$0.49 & $\pm$3.99 & $\pm$16.0    & $\pm$0.22 & $\pm$6.04 & $\pm$0.00 & $\pm$6.29 & $\pm$6.8  & $\pm$0.08 & $\pm$0.06 & $\pm$0.05 & $\pm$0.25 \\
M20150807\_014506 & 73.63     & 58.97     & 241.75       & 0.99      & 115.97    & 134.06    & 273.26    & $-$0.34   & 6.92      & $-$4.69   & $-$4.96   & 1.44 \\
(Canary)          & $\pm$0.19 & $\pm$0.14 & $\pm$2289.9  & $\pm$0.01 & $\pm$0.12 & $\pm$0.00 & $\pm$0.55 & $\pm$1.94 & $\pm$0.00 & $\pm$0.02 & $\pm$0.02 & $\pm$0.12 \\
M20150816\_005729 & 86.55     & 57.28     & 6.85         & 0.87      & 106.89    & 142.66    & 132.51    &    0.45   & 7.11      & $-$5.15   & $-$5.41   & 1.72 \\
(Canary)          & $\pm$0.49 & $\pm$0.66 & $\pm$2.13    & $\pm$0.03 & $\pm$1.2  & $\pm$0.00 & $\pm$1.99 & $\pm$1.28 & $\pm$0.01 & $\pm$0.02 & $\pm$0.02 & $\pm$0.15 \\
M20150816\_025513 & 90.28     & 60.5      & --           & 1.01      & 111.96    & 142.74    & 146.74    &      --   & 7.11      & $-$5.66   & $-$5.94   & 1.42 \\
(Canary)          & $\pm$1.22 & $\pm$1.2  &              & $\pm$0.12 & $\pm$0.48 & $\pm$0.00 & $\pm$2.24 &           & $\pm$0.02 & $\pm$0.03 & $\pm$0.02 & $\pm$0.18 \\
M20150816\_042120 & 82.79     & 60.87     & 11.5         & 0.92      & 116.01    & 142.8     & 146.27    & $-$0.04   & 7.15      & $-$5.08   & $-$5.36   & 1.01 \\
(Canary)          & $\pm$0.33 & $\pm$0.23 & $\pm$3.3     & $\pm$0.02 & $\pm$0.11 & $\pm$0.00 & $\pm$0.51 & $\pm$0.13 & $\pm$0.00 & $\pm$0.02 & $\pm$0.02 & $\pm$0.09 \\
M20150817\_044102 & 76.55     & 59.28     & 7.73         & 0.98      & 174.71    & 323.79    & 137.25    &    0.21   & 7.54      & $-$4.82   & $-$5.09   & 1.11 \\
(Canary)          & $\pm$0.26 & $\pm$0.24 & $\pm$1.5     & $\pm$0.00 & $\pm$0.51 & $\pm$0.00 & $\pm$1.07 & $\pm$0.09 & $\pm$0    & $\pm$0.02 & $\pm$0.02 & $\pm$0.11 \\
M20150824\_033620 & 87.55     & 59.92     & 26.47        & 0.97      & 114.8     & 150.47    & 240.51    & $-$0.24   & 7.47      & $-$5.59   & $-$5.87   & 1.01 \\
(Canary)          & $\pm$0.29 & $\pm$0.45 & $\pm$10434.5 & $\pm$0.03 & $\pm$0.35 & $\pm$0.00 & $\pm$1.47 & $\pm$56.1 & $\pm$0.01 & $\pm$0.02 & $\pm$0.02 & $\pm$0.11 \\
M20150826\_044420 & 92.84     & 64.47     & 7.13         & 0.86      & 129.14    & 332.45    & 350.99    & $-$0.04   & 6.49      & $-$5.72   & $-$6.03   & 1.3 \\
(Canary)          & $\pm$0.22 & $\pm$0.44 & $\pm$3.1     & $\pm$0.04 & $\pm$0.21 & $\pm$0.00 & $\pm$0.37 & $\pm$0.22 & $\pm$0.01 & $\pm$0.02 & $\pm$0.01 & $\pm$0.11 \\
M20150903\_033450 & 81.61     & 72.39     & --           & 1.01      & 176.91    & 340.15    & 337.58    &      --   & 7         & $-$4.96   & $-$5.32   & 2.74 \\
(Canary)          & $\pm$0.25 & $\pm$0.34 &              & $\pm$0.03 & $\pm$0.45 & $\pm$0.00 & $\pm$1.59 &           & $\pm$0.01 & $\pm$0.02 & $\pm$0.02 & $\pm$0.25 \\
M20150905\_042433 & 78.04     & 60.41     & 7.64         & 0.97      & 142.99    & 162.1     &  55.53    &    0.21   & 7.05      & $-$4.62   & $-$4.9    & 3.68 \\
(Canary)          & $\pm$0.54 & $\pm$0.57 & $\pm$38.51   & $\pm$0.01 & $\pm$0.66 & $\pm$0.00 & $\pm$1.71 & $\pm$1.57 & $\pm$0.01 & $\pm$0.01 & $\pm$0.01 & $\pm$0.25 \\
M20150909\_035937 & 78.94     & 64.41     & 26.16        & 0.96      & 126.74    & 165.96    & 210.68    & $-$0.5    & 6.43      & $-$4.93   & $-$5.24   & 1.3 \\
(Canary)          & $\pm$0.31 & $\pm$0.28 & $\pm$413.19  & $\pm$0.02 & $\pm$0.37 & $\pm$0.00 & $\pm$0.54 & $\pm$3.76 & $\pm$0.00 & $\pm$0.02 & $\pm$0.02 & $\pm$0.11 \\
M20150911\_055633 & 75.35     & 61.7      & --           & 1.02      & 113.82    & 347.99    & 352.53    &      --   & 8.46      & $-$4.53   & $-$4.82   & 1.22 \\
(Canary)          & $\pm$0.16 & $\pm$0.26 &              & $\pm$0.03 & $\pm$0.25 & $\pm$0.00 & $\pm$0.33 &           & $\pm$0.00 & $\pm$0.02 & $\pm$0.02 & $\pm$0.17 \\
M20150915\_052641 & 93.95     & 61.46     & 14.5         & 0.93      & 115.62    & 351.86    &   7.72    & $-$0.18   & 6.84      & $-$5.97   & $-$6.26   & 1.95 \\
(Canary)          & $\pm$1.03 & $\pm$0.64 & $\pm$893.85  & $\pm$0.07 & $\pm$0.33 & $\pm$0.00 & $\pm$4.64 & $\pm$6.7  & $\pm$0.01 & $\pm$0.02 & $\pm$0.02 & $\pm$0.2 \\
M20150927\_224230 & 91.16     & 64.43     & 3.98         & 0.83      & 142.62    & 184.29    & 251.34    &    0.52   & 7.22      & $-$5.06   & $-$5.37   & 1.91 \\
(Canary)          & $\pm$0.37 & $\pm$0.57 & $\pm$0.87    & $\pm$0.03 & $\pm$0.31 & $\pm$0.00 & $\pm$1.72 & $\pm$0.24 & $\pm$0.01 & $\pm$0.02 & $\pm$0.02 & $\pm$0.22 \\
M20150930\_020632 & 80.62     & 63.39     & 7.54         & 0.93      & 138.75    & 186.39    & 267.68    &    0.02   & 7.2       & $-$5.25   & $-$5.55   & 4.96 \\
(Canary)          & $\pm$1.00 & $\pm$2.09 & $\pm$48.8    & $\pm$0.04 & $\pm$7.17 & $\pm$0.00 & $\pm$9.17 & $\pm$4.39 & $\pm$0.04 & $\pm$0.07 & $\pm$0.06 & $\pm$1.6 \\
\hline
\end{tabular}
\end{table*}
\begin{table*}
\setcounter{table}{2}
\caption{Properties of selected cometary meteors on cometary orbits from two
  databases --- continued from previous page.}%
\vspace*{6pt}
\centering
\begin{tabular}{@{}c@{\0}c@{\0}c@{\0}c@{\0}c@{\0}c@{\0}c@{\0}c@{\0}c@{\0}c@{\0}c@{\0}c@{\0}c@{}}
\hline
Meteor name & $h_\mathrm{end}$ & $v_\mathrm{beg}$ & $a$  & $e$       & $i$       & $\Omega$  & $\omega$  & $T_J$  & $K_B$ & $P_E$ & $P_{E,\mathrm{mod}}$ & $m_\mathrm{phot}$ \\
and database      &  [km]     & [km/s]    & [au]         &           & [ \g\ ]   & [ \g\ ]   & [ \g\ ]   &           &           &           &           & [g] \\
\hline
M20151007\_231646 & 88.66     & 68.25     & --           & 1.09      & 146.87    & 194.15    & 250.99    &      --   & 7.1       & $-$5.05   & $-$5.39   & 1.57 \\
(Canary)          & $\pm$0.47 & $\pm$0.29 &              & $\pm$0.02 & $\pm$0.18 & $\pm$0.00 & $\pm$0.91 &           & $\pm$0.00 & $\pm$0.01 & $\pm$0.01 & $\pm$0.08 \\
M20151013\_042644 & 79.62     & 61.43     & 2.91         & 0.7       & 126.02    &  19.31    &  45.58    &    1.16   & 7.82      & $-$4.86   & $-$5.15   & 1.02 \\
(Canary)          & $\pm$0.35 & $\pm$0.24 & $\pm$0.19    & $\pm$0.02 & $\pm$0.38 & $\pm$0.00 & $\pm$0.78 & $\pm$0.33 & $\pm$0.00 & $\pm$0.03 & $\pm$0.03 & $\pm$0.14 \\
M20151105\_061959 & 81.92     & 69.7      & --           & 1.05      & 155.33    & 222.3     & 111.45    &      --   & 6.07      & $-$5.21   & $-$5.55   & 3.55 \\
(Canary)          & $\pm$0.37 & $\pm$1.46 &              & $\pm$0.1  & $\pm$0.42 & $\pm$0.00 & $\pm$3.65 &           & $\pm$0.02 & $\pm$0.03 & $\pm$0.02 & $\pm$0.46 \\
M20151106\_035602 & 89.59     & 68.62     & 10.64        & 0.94      & 164.12    &  43.19    &  71.53    & $-$0.46   & 6.87      & $-$5.73   & $-$6.06   & 2.04 \\
(Canary)          & $\pm$3.39 & $\pm$0.7  & $\pm$31.72   & $\pm$0.04 & $\pm$1.26 & $\pm$0.00 & $\pm$4.85 & $\pm$0.45 & $\pm$0.01 & $\pm$0.04 & $\pm$0.04 & $\pm$0.41 \\
M20151112\_043611 & 80.01     & 60        & 38.56        & 0.99      & 115.81    &  49.25    &  84.47    & $-$0.25   & 6.27      & $-$4.97   & $-$5.25   & 1.11 \\
(Canary)          & $\pm$0.11 & $\pm$0.21 & $\pm$1617    & $\pm$0.01 & $\pm$0.2  & $\pm$0.00 & $\pm$0.55 & $\pm$5.87 & $\pm$0.00 & $\pm$0.02 & $\pm$0.02 & $\pm$0.11 \\
M20151117\_005257 & 87.89     & 68.11     & 9.99         & 0.91      & 141.21    & 234.13    & 213       & $-$0.39   & 6.6       & $-$4.98   & $-$5.32   & 3.89 \\
(Canary)          & $\pm$0.48 & $\pm$0.37 & $\pm$3.11    & $\pm$0.03 & $\pm$0.33 & $\pm$0.00 & $\pm$0.87 & $\pm$0.28 & $\pm$0.01 & $\pm$0.05 & $\pm$0.05 & $\pm$1.1 \\
M20151117\_031022 & 82.73     & 56.97     & --           & 1.04      & 102.44    &  54.23    & 106.36    &      --   & 6.91      & $-$5.2    & $-$5.45   & 1.22 \\
(Canary)          & $\pm$0.4  & $\pm$0.4  &              & $\pm$0.02 & $\pm$1.26 & $\pm$0.00 & $\pm$1.61 &           & $\pm$0.01 & $\pm$0.02 & $\pm$0.02 & $\pm$0.14 \\
M20151118\_051249 & 91.26     & 56.8      & --           & 1.01      & 104.59    & 235.32    &  57.32    &      --   & 6.86      & $-$5.25   & $-$5.5    & 1.35 \\
(Canary)          & $\pm$0.2  & $\pm$0.15 &              & $\pm$0.00 & $\pm$0.34 & $\pm$0.00 & $\pm$0.33 &           & $\pm$0.00 & $\pm$0.01 & $\pm$0.01 & $\pm$0.08 \\
M20151118\_053828 & 96.07     & 55.04     & 5.78         & 0.83      &  98.24    & 235.34    & 198.08    &    0.71   & 6.82      & $-$6.28   & $-$6.52   & 1.52 \\
(Canary)          & $\pm$0.2  & $\pm$0.11 & $\pm$0.36    & $\pm$0.01 & $\pm$0.11 & $\pm$0.00 & $\pm$0.3  & $\pm$0.13 & $\pm$0.00 & $\pm$0.02 & $\pm$0.02 & $\pm$0.15 \\
M20151121\_020703 & 79.53     & 61.6      & --           & 1.06      & 117.14    & 238.22    & 263.78    &      --   & 7.09      & $-$5.01   & $-$5.3    & 2.75 \\
(Canary)          & $\pm$0.47 & $\pm$0.37 &              & $\pm$0.01 & $\pm$0.51 & $\pm$0.00 & $\pm$1.71 &           & $\pm$0.01 & $\pm$0.03 & $\pm$0.03 & $\pm$0.49 \\
M20151127\_021401 & 83.73     & 53.83     & --           & 1.02      &  90.89    &  64.28    &  56.73    &      --   & 7         & $-$5.14   & $-$5.37   & 2.45 \\
(Canary)          & $\pm$0.4  & $\pm$0.59 &              & $\pm$0.03 & $\pm$0.73 & $\pm$0.00 & $\pm$0.76 &           & $\pm$0.01 & $\pm$0.04 & $\pm$0.04 & $\pm$0.48 \\
M20151130\_024431 & 93.68     & 56.55     & 3.00         & 0.71      & 106.75    &  67.34    &  45.28    &    1.45   & 6.94      & $-$5.78   & $-$6.03   & 2.06 \\
(Canary)          & $\pm$0.88 & $\pm$0.94 & $\pm$0.53    & $\pm$0.04 & $\pm$1.05 & $\pm$0.00 & $\pm$4.89 & $\pm$1.07 & $\pm$0.02 & $\pm$0.03 & $\pm$0.03 & $\pm$0.28 \\
M20151212\_031019 & 79.72     & 61.2      & 3.21         & 0.69      & 118.92    & 259.54    & 178.9     &    1.06   & 6.93      & $-$5.02   & $-$5.3    & 10.5 \\
(Canary)          & $\pm$0.73 & $\pm$0.65 & $\pm$0.84    & $\pm$0.06 & $\pm$0.35 & $\pm$0.00 & $\pm$0.33 & $\pm$0.37 & $\pm$0.01 & $\pm$0.02 & $\pm$0.02 & $\pm$1.08 \\
M20151215\_052144 & 86.48     & 53.45     & 20.25        & 0.97      &  90.32    & 262.68    & 108.21    &    0.23   & 7.09      & $-$5.26   & $-$5.49   & 2.05 \\
(Canary)          & $\pm$0.25 & $\pm$0.26 & $\pm$16.48   & $\pm$0.01 & $\pm$0.36 & $\pm$0    & $\pm$0.43 & $\pm$0.36 & $\pm$0.01 & $\pm$0.02 & $\pm$0.02 & $\pm$0.17 \\
M20151219\_001822 & 87.85     & 67.67     & 14.19        & 0.97      & 178       & 266.59    & 273.97    & $-$0.47   & 6.62      & $-$4.8    & $-$5.13   & 1.21 \\
(Canary)          & $\pm$0.64 & $\pm$0.49 & $\pm$182.82  & $\pm$0.02 & $\pm$0.33 & $\pm$0.01 & $\pm$2.02 & $\pm$0.73 & $\pm$0.01 & $\pm$0.02 & $\pm$0.01 & $\pm$0.09 \\
M20160325\_074148 & 90.18     & 59.1      & 5.65         & 0.83      & 110.89    & 184.87    & 336.37    &    0.51   & 6.97      & $-$5.54   & $-$5.81   & 1.17 \\
(Chile)           & $\pm$0.23 & $\pm$0.2  & $\pm$0.55    & $\pm$0.02 & $\pm$0.25 & $\pm$0.00 & $\pm$0.32 & $\pm$0.27 & $\pm$0.00 & $\pm$0.02 & $\pm$0.02 & $\pm$0.14 \\
M20160517\_095531 & 82.23     & 66.88     & 7.99         & 0.93      & 171.65    &  56.69    & 262.34    & $-$0.27   & 7.74      & $-$5.62   & $-$5.95   & 3.85 \\
(Chile)           & $\pm$0.4  & $\pm$0.68 & $\pm$18.7    & $\pm$0.09 & $\pm$3.98 & $\pm$0.09 & $\pm$5.87 & $\pm$0.88 & $\pm$0.01 & $\pm$0.05 & $\pm$0.05 & $\pm$1.00 \\
M20160605\_021236 & 103.91    & 58.26     & --           & 1.04      & 107.33    &  74.63    & 269.56    &      --   & 7.23      & $-$5.62   & $-$5.89   & 2.54 \\
(Chile)           & $\pm$0.32 & $\pm$0.19 &              & $\pm$0.01 & $\pm$0.28 & $\pm$0.00 & $\pm$0.42 &           & $\pm$0.00 & $\pm$0.02 & $\pm$0.02 & $\pm$0.18 \\
M20160618\_084448 & 88.46     & 56.58     & 16.78        & 0.94      & 101.93    & 267.32    & 352.56    &    0.03   & 7.26      & $-$5.8    & $-$6.05   & 1.46 \\
(Chile)           & $\pm$0.24 & $\pm$0.37 & $\pm$66.28   & $\pm$0.05 & $\pm$0.62 & $\pm$0.00 & $\pm$2.59 & $\pm$0.79 & $\pm$0.01 & $\pm$0.04 & $\pm$0.04 & $\pm$0.31 \\
M20160620\_095604 & 84.59     & 56.48     & --           & 1.00      & 106.98    & 269.28    & 253.48    &      --   & 6.6       & $-$5.36   & $-$5.61   & 1.00 \\
(Chile)           & $\pm$0.09 & $\pm$0.12 &              & $\pm$0.00 & $\pm$0.18 & $\pm$0.00 & $\pm$0.33 &           & $\pm$0.00 & $\pm$0.02 & $\pm$0.02 & $\pm$0.11 \\
M20160719\_040146 & 75.6      & 59.46     & --           & 1.01      & 157.77    & 296.7     & 142       &      --   & 7.97      & $-$4.88   & $-$5.16   & 2.63 \\
(Chile)           & $\pm$0.16 & $\pm$0.46 &              & $\pm$0.00 & $\pm$1.38 & $\pm$0.00 & $\pm$1.12 &           & $\pm$0.01 & $\pm$0.03 & $\pm$0.03 & $\pm$0.47 \\
M20160727\_092303 & 86.08     & 66.95     & 2.55         & 0.65      & 178.88    & 304.57    &  44.11    &    0.97   & 7.33      & $-$5.9    & $-$6.22   & 5.78 \\
(Chile)           & $\pm$0.49 & $\pm$0.79 & $\pm$0.41    & $\pm$0.04 & $\pm$1.43 & $\pm$0.1  & $\pm$5.99 & $\pm$0.25 & $\pm$0.01 & $\pm$0.03 & $\pm$0.03 & $\pm$0.81 \\
M20160820\_092005 & 73.27     & 59.18     & --           & 1.01      & 142.04    & 327.56    & 134.42    &      --   & 7.08      & $-$5.00   & $-$5.27   & 3.85 \\
(Chile)           & $\pm$0.15 & $\pm$0.35 &              & $\pm$0.01 & $\pm$0.7  & $\pm$0.00 & $\pm$0.83 &           & $\pm$0.01 & $\pm$0.03 & $\pm$0.03 & $\pm$0.55 \\
M20160825\_042750 & 86.13     & 50.73     & 224.32       & 0.99      &  85.55    & 152.18    & 239.3     &    0.12   & 7.28      & $-$4.96   & $-$5.17   & 1.23 \\
(Chile)           & $\pm$0.27 & $\pm$0.18 & $\pm$457.71  & $\pm$0.01 & $\pm$0.23 & $\pm$0.00 & $\pm$0.33 & $\pm$0.29 & $\pm$0.00 & $\pm$0.01 & $\pm$0.01 & $\pm$0.09 \\
M20160925\_072409 & 77.07     & 63.89     & 17.57        & 0.98      & 166.66    & 182.44    & 291.11    & $-$0.39   & 7.54      & $-$4.88   & $-$5.18   & 2.98 \\
(Chile)           & $\pm$0.3  & $\pm$0.94 & $\pm$282.3   & $\pm$0.02 & $\pm$0.23 & $\pm$0.00 & $\pm$4.02 & $\pm$0.96 & $\pm$0.02 & $\pm$0.02 & $\pm$0.02 & $\pm$0.35 \\
M20161105\_073719 & 99.39     & 67.68     & --           & 1.07      & 130.36    & 223.1     & 186.96    &      --   & 6.99      & $-$5.42   & $-$5.75   & 1.93 \\
(Chile)           & $\pm$0.07 & $\pm$0.11 &              & $\pm$0.01 & $\pm$0.09 & $\pm$0.00 & $\pm$0.04 &           & $\pm$0.00 & $\pm$0.01 & $\pm$0.01 & $\pm$0.13 \\
\hline
\end{tabular}
\end{table*}

\begin{figure*}[!t]
\centering
\includegraphics[width = \textwidth]{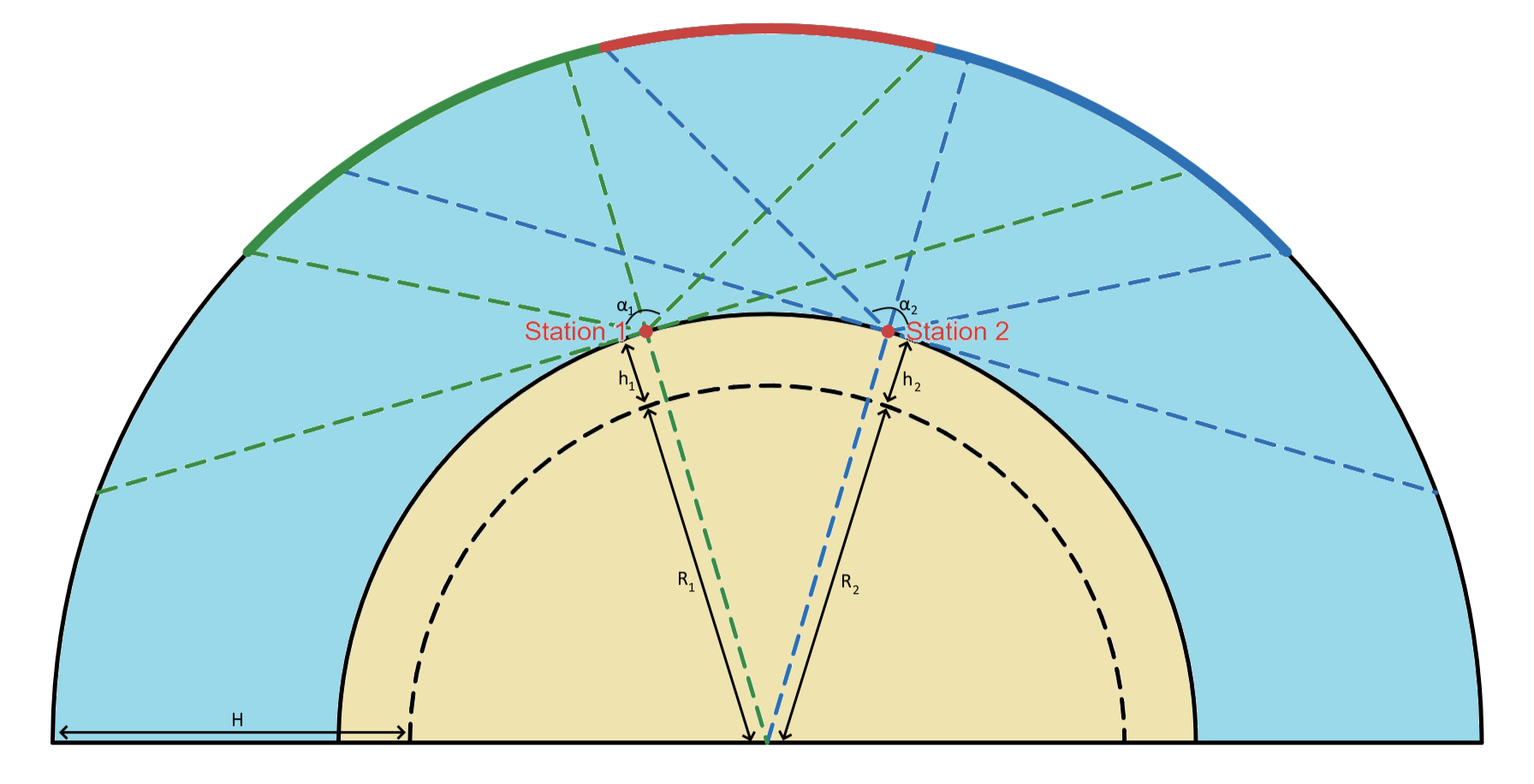}%
\caption{Representation of the effective area for a double-station meteor
  observation (schematic side view). Outer, light blue area represents the
  atmosphere, inner yellow area represents the Earth, dashed black line is
  the sea level. $H$ is the height above sea level for which the effective
  area is calculated, $R_1, R_2$ are the Earth's radii at station locations,
  $h_1, h_2$ are altitudes of each station, angles $\alpha_1, \alpha_2$
  represent angular fields of view of each camera. Green (towards the left)
  and dark blue (towards the right) areas are the areas seen by individual
  stations. Red area in middle, where green and dark blue overlap, is the
  effective area $A_\mathrm{eff}$ observed by both cameras.}
\label{fig:side}
\end{figure*}

\section{Flux determination}

Meteor flux is defined as a number of meteors passing through a unit of area
in a unit of time:

\begin{equation}
F(m>m_\mathrm{lim}) =
  \frac {N(m>m_\mathrm{lim})} {A_\mathrm{eff} \cdot t_\mathrm{obs}}
\label{eqn:flux}
\end{equation}

where $N(m>m_\mathrm{lim})$ is the number of meteoroids with mass larger than
$m_\mathrm{lim}$, $A_\mathrm{eff}$ is the effective area of the atmosphere
observed by the system at given height in the atmosphere, $t_\mathrm{obs}$ is
the total observational time \cite{vida22}.

\subsection*{Observational time}

The observational time is affected by various effects such as weather
conditions (mostly clouds), lunar phase (only bright meteors are visible
during the full Moon), seasonal variations of night duration, technical and
software maintenance, etc. The AMOS cameras perform observations every night
(including partly cloudy nights) between nautical twilights (the Sun is 12\g\
below the horizon), independently of weather conditions and Moon phase.
Cloudy nights at the locations of the AMOS stations are rare in Chile and
seasonal in the Canary Islands (primarily during the winter months). Lunar
phase was not problematic since we were focusing on bright meteors.

For double-station observations (as we have in our databases), the total
observational time is reduced because we need to estimate the total time when
two cameras observed the same meteor simultaneously. For this, we used two
approaches. In the first approach, times of observations were taken directly
from the databases for each night as a period between the first and the last
observed double-station meteor with one hour precision. Only nights with 3
and more detected meteors were considered (correction for weather). The
Canary database contains data collected between 2015 June 1 and 2015 December
26, 208 nights in total. For the Chile database, there were 248 nights of
observations (from 2016 March 21 to 2016 November 24). Removing maintenance
periods and nights with less than 3 meteors, we obtained 139 nights of
observations for Canary stations and 176 nights for Chile stations. The
estimated observational time for the Canary database is 911 hours, for the
Chile database is 1314 hours.

In the second approach, we again selected nights with 3 and more meteors. We
did it for each month and then multiplied the number of nights by the mean
duration of the night for each particular month (taken as the period between
nautical twilights of the night in the middle of the particular month for the
geographical location of the AMOS station). These times were corrected for
the weather and for waiting for a common meteor by subtracting 2 hours from
each mean night duration. This approach yielded similar results to the first
approach: 1070 hours for the Canary database and 1568.5 hours for the Chile
database. Our expert evaluation of observational time error is
$\sim$10--20\%. For time-area products, we took mean values from these two
approaches.

\subsection*{AMOS effective area}

For a single-station observation, the effective area is a 3D surface area of
the atmosphere observed by the camera at a given height corrected for
observational biases. It depends on the camera FOV, station geographical
location and altitude. Biases include various natural phenomena, such as
weather conditions, atmospheric extinction (especially for large zenith
distances), obstacles in the FOV, and instrumental effects, e.g.\ sensitivity
of the camera, detection software sensitivity, vignetting (drop of brightness
at the periphery of the image compared to its center), imperfections in the
optical system, etc.\ \cite{vida22}. The effective area is always smaller than
the true area, because biases decrease the number of observed meteors
compared to ideal conditions. For multiple-station observations, the total
effective area is given by the intersection of effective areas of individual
stations (Figure~\ref{fig:side}). There are several methods of calculating
$A_\mathrm{eff}$, for example, debiasing method based on the identification
and compensation of observational biases \cite{vida22}, statistical Monte
Carlo approach involving parametric simulation of meteor data and matching
them with observations \cite{balaz20}, dividing the area into small cells
with known areas and calculating the number of cells visible by multiple
stations \cite{halliday96}, or determining coordinates of the farthest
observed meteors from a large set of observations \cite{borovicka22I}.

For our purposes, we decided to follow a simplified geometrical approach.
Parameters for the real AMOS stations used in our analysis are provided in
Table~\ref{tab:coord}. First, we needed to determine the area seen by a
single camera. For simplicity, we neglected the differences between station
altitudes (cameras in the Canary Islands and Chile are located at
approximately equal heights of $\sim$2.4 km above the sea level). Parameters
of the AMOS San Pedro station were taken as reference. The area was
calculated for the altitude of $H$ = 70 km above sea level, following
Halliday et al.\ (1996) and Vida et al.\ (2023). \nocite{halliday96,vida23}
The AMOS cameras have a FOV of 180\g$\times$140\g. If there were no biases,
we would see meteors (which are 70 km above the ground) on the local horizon
at a maximum distance of $\sim$931 km from the station. However, this value
is too optimistic, since close to the horizon meteors are hardly detectable
due to atmospheric extinction and vignetting which decrease visual
brightness.  As a correction for these effects, we set a lower limit for the
altitude above the horizon to 3\g, corresponding to $\alpha$ = 174\g\
(Figure~\ref{fig:side}) and maximum distance of $\sim$655 km at which our
cameras would see bright meteors. Since we did not know exactly how $\alpha$
changed with azimuth around the station, we used the GeoGebra 3D
Calculator\footnote{\texttt{https://www.geogebra.org}} to determine
coordinates of edges of the field of view of the camera. We did it by finding
the intersection of the spherical surface of radius $R + H$ (representing the
surface in the atmosphere at altitude $H$) with an infinite cone with an
aperture angle of 174\g\ originating in the camera location cut by two
infinite planes passing through the station with the angle 140\g\ between
them. The intersection was found using the function IntersectPath().  This
way we obtained 3D coordinates of edges of the surface area seen by the
camera with the FOV of 180\g$\times$140\g, with the area below 3\g\ altitude
neglected. To calculate the single-station effective area for our reference
station, we used a 2D projection approximation. We estimated the area using
the Measure tool in GeoGebra, the result is $\sim$392000 km$^2$.

\begin{figure}[htb]
\centering
\includegraphics[width = \columnwidth]{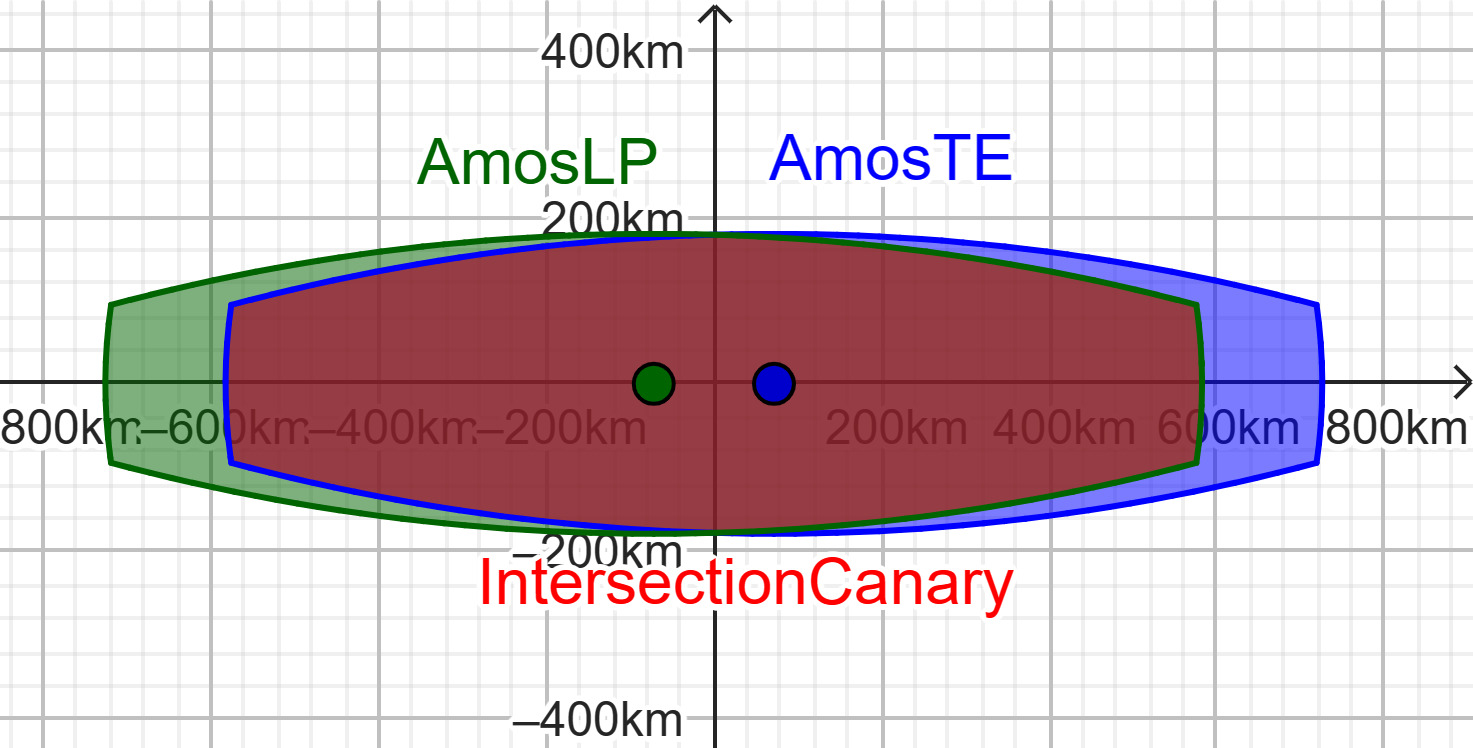}%
\caption{Representation of the effective area of the Canary stations
  (schematic top view). Green and blue dots represent individual cameras.
  Stations are aligned along x-axis. Green (left) and blue (right) areas are
  the areas seen by individual stations. For each camera, the longer side of
  the green/blue polygon corresponds to the longer side of the FOV. Central
  red intersection is the effective area.}
\label{fig:top}
\end{figure}

To estimate the effective area seen by two cameras, we first determined the
distances between individual stations (Figure~\ref{fig:map}), using the known
geographical coordinates and altitudes of the AMOS stations
(Table~\ref{tab:coord}) and Google Earth
Pro\footnote{\texttt{https://earth.google.com/intl/earth/versions/\#earth-pro}}
software. The AMOS cameras are oriented in such a way that the longer sides
of the FOV are aligned with the line connecting two stations. We used
distances between stations as separations between 2D projections of
single-station effective areas (neglecting Earth's curvature). The
double-station effective area was calculated as the area of intersection of
two single-station 2D polygons using the IntersectPath() and Measure tools in
GeoGebra (Figure~\ref{fig:top}). The resulting effective area for the AMOS
Canary database is 341000 km$^2$, for the Chile database is 363000 km$^2$.
Our expert evaluation of the effective area error is $\sim$10--20\%, mainly
due to atmospheric extinction and vignetting. The corresponding time-area
products are 337760500 km$^2$$\cdot$h for the Canary database and 523173750
km$^2$$\cdot$h for the Chile database. Obtaining these values was the last
step to calculate meteor fluxes using (\ref{eqn:flux}).

\section{Conclusions and future work}

In this work we presented our motivation and methodology of looking for
asteroidal meteoroids coming from the Oort cloud. The analysis of two AMOS
databases yielded 9 rocky meteoroids on cometary orbits, compared with 53
cometary meteoroids found on similar orbits. We also estimated time-area
products for two analyzed databases. These estimates will be used to
calculate meteor fluxes, from which we will estimate the ice-to-rock ratio
and compare it with predictions of different cosmogonic models. The results
will be published in a separate paper.

\section*{Acknowledgement}

We would like to express gratitude to the AMOS Team for the provided data,
and to Dr.\ Denis Vida, who kindly shared his code to estimate the
uncertainties of the final results.

%\nocite{*}
\bibliographystyle{imo2}
\bibliography{kuksenko-cosmogonic}

@ARTICLE{balaz20,
       author = {{Bal{\'a}{\v{z}}}, Martin and {T{\'o}th}, Juraj and {Vere{\v{s}}}, Peter and {Jedicke}, Robert},
        title = "{ASMODEUS meteor simulation tool}",
      journal = {Planetary and Space Science},
         year = 2020,
        month = oct,
       volume = {190},
          eid = {104937},
        pages = {104937},
          doi = {10.1016/j.pss.2020.104937},
       adsurl = {https://ui.adsabs.harvard.edu/abs/2020P&SS..19004937B}
}

@ARTICLE{borovicka22I,
       author = {{Borovi\v{c}ka}, J. and {Spurn\'y}, P. and {Shrben\'y}, L. and {\v{S}tork}, R. and {Kotkov\'a}, L. and {Fuchs}, J. and {Kecl\'ikov\'a}, J. and {Zichov\'a}, H. and {M\'anek}, J. and {V\'achov\'a}, P. and Macourkov\'a, I. and Svore\v{n}, J. and Mucke, H.},
        title = "{Data on 824 fireballs observed by the digital cameras of the European Fireball Network in 2017--2018. I. Description of the network, data reduction procedures, and the catalog}",
      journal = {Astronomy and Astrophysics},
         year = 2022,
        month = nov,
       volume = {667},
          eid = {A157},
        pages = {A157},
          doi = {10.1051/0004-6361/202244184},
archivePrefix = {arXiv},
       eprint = {2209.11186},
 primaryClass = {astro-ph.EP},
       adsurl = {https://ui.adsabs.harvard.edu/abs/2022A&A...667A.157B}
}

@ARTICLE{borovicka22II,
       author = {{Borovi\v{c}ka}, J. and {Spurn\'y}, P. and {Shrben\'y}, L.},
        title = "{Data on 824 fireballs observed by the digital cameras of the European Fireball Network in 2017--2018. II. Analysis of orbital and physical properties of centimeter-sized meteoroids}",
      journal = {Astronomy and Astrophysics},
         year = 2022,
        month = nov,
       volume = {667},
          eid = {A158},
        pages = {A158},
          doi = {10.1051/0004-6361/202244197},
archivePrefix = {arXiv},
       eprint = {2209.11254},
 primaryClass = {astro-ph.EP},
       adsurl = {https://ui.adsabs.harvard.edu/abs/2022A&A...667A.158B}
}

@ARTICLE{ceplecha58,
       author = {{Ceplecha}, Z.},
        title = "{On the composition of meteors}",
      journal = {Bulletin of the Astronomical Institutes of Czechoslovakia},
         year = 1958,
        month = jan,
       volume = {9},
        pages = {154-159},
       adsurl = {https://ui.adsabs.harvard.edu/abs/1958BAICz...9..154C}
}

@ARTICLE{cm76,
       author = {{Ceplecha}, Z. and {McCrosky}, R.~E.},
        title = "{Fireball end heights: A diagnostic for the structure of meteoric material}",
      journal = {Journal of Geophysical Research},
         year = 1976,
        month = dec,
       volume = {81},
       number = {B35},
        pages = {6257-6275},
          doi = {10.1029/JB081i035p06257},
       adsurl = {https://ui.adsabs.harvard.edu/abs/1976JGR....81.6257C}
}

@ARTICLE{halliday96,
       author = {{Halliday}, Ian and {Griffin}, Arthur A. and {Blackwell}, Alan T.},
        title = "{Detailed data for 259 fireballs from the Canadian camera network and inferences concerning the influx of large meteoroids}",
      journal = {Meteoritics and Planetary Science},
         year = 1996,
        month = mar,
       volume = {31},
        pages = {185-217},
          doi = {10.1111/j.1945-5100.1996.tb02014.x},
       adsurl = {https://ui.adsabs.harvard.edu/abs/1996M&PS...31..185H}
}

@ARTICLE{kresak69,
       author = {{Kres{\'a}k}, L'.},
        title = "{The discrimination between cometary and asteroidal meteors. I. The orbital criteria}",
      journal = {Bulletin of the Astronomical Institutes of Czechoslovakia},
         year = 1969,
        month = jan,
       volume = {20},
        pages = {177-188},
       adsurl = {https://ui.adsabs.harvard.edu/abs/1969BAICz..20..177K}
}

@ARTICLE{lj12,
       author = {{Lambrechts}, M. and {Johansen}, A.},
        title = "{Rapid growth of gas-giant cores by pebble accretion}",
      journal = {Astronomy and Astrophysics},
         year = 2012,
        month = aug,
       volume = {544},
          eid = {A32},
        pages = {A32},
          doi = {10.1051/0004-6361/201219127},
archivePrefix = {arXiv},
       eprint = {1205.3030},
 primaryClass = {astro-ph.EP},
       adsurl = {https://ui.adsabs.harvard.edu/abs/2012A&A...544A..32L}
}

@INPROCEEDINGS{levison96,
       author = {{Levison}, H.~F.},
        title = "{Comet taxonomy}",
    booktitle = {Completing the Inventory of the Solar System},
         year = 1996,
       editor = {{Rettig}, Terrence and {Hahn}, Joseph M.},
       series = {ASP Conference Series},
       volume = {107},
   publisher = {Astronomical Society of the Pacific},
        month = jan,
        pages = {173-191},
       adsurl = {https://ui.adsabs.harvard.edu/abs/1996ASPC..107..173L}
}

@ARTICLE{meech16,
       author = {{Meech}, Karen J. and {Yang}, Bin and {Kleyna}, Jan and {Hainaut}, Olivier R. and {Berdyugina}, Svetlana and {Keane}, Jacqueline V. and {Micheli}, Marco and {Morbidelli}, Alessandro and {Wainscoat}, Richard J.},
        title = "{Inner solar system material discovered in the Oort cloud}",
      journal = {Science Advances},
         year = 2016,
        month = apr,
       volume = {2},
          eid = {e1600038},
        pages = {e1600038},
          doi = {10.1126/sciadv.1600038},
       adsurl = {https://ui.adsabs.harvard.edu/abs/2016SciA....2E0038M}
}

@ARTICLE{pc83,
       author = {{Pecina}, P. and {Ceplecha}, Z.},
        title = "{New aspects in single-body meteor physics}",
      journal = {Bulletin of the Astronomical Institutes of Czechoslovakia},
         year = 1983,
        month = mar,
       volume = {34},
        pages = {102-121},
       adsurl = {https://ui.adsabs.harvard.edu/abs/1983BAICz..34..102P}
}

@ARTICLE{picone02,
       author = {{Picone}, J.~M. and {Hedin}, A.~E. and {Drob}, D.~P. and {Aikin}, A.~C.},
        title = "{NRLMSISE-00 empirical model of the atmosphere: Statistical comparisons and scientific issues}",
      journal = {Journal of Geophysical Research (Space Physics)},
         year = 2002,
        month = dec,
       volume = {107},
       number = {A12},
          eid = {1468},
        pages = {1468},
          doi = {10.1029/2002JA009430},
       adsurl = {https://ui.adsabs.harvard.edu/abs/2002JGRA..107.1468P}
}

@ARTICLE{raymond09,
       author = {{Raymond}, Sean N. and {O'Brien}, David P. and {Morbidelli}, Alessandro and {Kaib}, Nathan A.},
        title = "{Building the terrestrial planets: Constrained accretion in the inner Solar System}",
      journal = {Icarus},
         year = 2009,
        month = oct,
       volume = {203},
        pages = {644-662},
          doi = {10.1016/j.icarus.2009.05.016},
archivePrefix = {arXiv},
       eprint = {0905.3750},
 primaryClass = {astro-ph.EP},
       adsurl = {https://ui.adsabs.harvard.edu/abs/2009Icar..203..644R}
}

@ARTICLE{shannon15,
       author = {{Shannon}, Andrew and {Jackson}, Alan P. and {Veras}, Dimitri and {Wyatt}, Mark},
        title = "{Eight billion asteroids in the Oort cloud}",
      journal = {Monthly Notices of the Royal Astronomical Society},
         year = 2015,
        month = jan,
       volume = {446},
        pages = {2059-2064},
          doi = {10.1093/mnras/stu2267},
archivePrefix = {arXiv},
       eprint = {1410.7403},
 primaryClass = {astro-ph.EP},
       adsurl = {https://ui.adsabs.harvard.edu/abs/2015MNRAS.446.2059S}
}

@ARTICLE{shannon19,
       author = {{Shannon}, Andrew and {Jackson}, Alan P. and {Wyatt}, Mark C.},
        title = "{Oort cloud asteroids: collisional evolution, the Nice Model, and the Grand Tack}",
      journal = {Monthly Notices of the Royal Astronomical Society},
         year = 2019,
        month = jun,
       volume = {485},
        pages = {5511-5518},
          doi = {10.1093/mnras/stz776},
archivePrefix = {arXiv},
       eprint = {1903.03199},
 primaryClass = {astro-ph.EP},
       adsurl = {https://ui.adsabs.harvard.edu/abs/2019MNRAS.485.5511S}
}

@INPROCEEDINGS{sb99,
       author = {{Spurn{\'y}}, P. and {Borovi{\v{c}}ka}, J.},
        title = "{EN010697 Karl{\v{s}}tejn: the first type I fireball on retrograde orbit}",
    booktitle = {Meteoroids 1998},
         year = 1999,
       editor = {{Baggaley}, W.~J. and {Porub\v{c}an}, V.},
   publisher = {Astronomical Institute, Slovak Academy of Sciences},
        month = jan,
        pages = {143-148},
       adsurl = {https://ui.adsabs.harvard.edu/abs/1999md98.conf..143S}
}

@ARTICLE{toth11,
       author = {{T{\'o}th}, Juraj and {Korno{\v{s}}}, Leonard and {Vere{\v{s}}}, Peter and {{\v{S}}ilha}, Jir{\'\i} and {Kalman{\v{c}}ok}, Du{\v{s}}an and {Zigo}, Pavol and {Vil{\'a}gi}, Jozef},
        title = "{All-sky video orbits of Lyrids 2009}",
      journal = {Publications of the Astronomical Society of Japan},
         year = 2011,
        month = apr,
       volume = {63},
        pages = {331-334},
          doi = {10.1093/pasj/63.2.331},
archivePrefix = {arXiv},
       eprint = {1106.0590},
 primaryClass = {astro-ph.EP},
       adsurl = {https://ui.adsabs.harvard.edu/abs/2011PASJ...63..331T}
}

@ARTICLE{toth15,
       author = {{T{\'o}th}, Juraj and {Korno{\v{s}}}, Leonard and {Zigo}, Pavol and {Gajdo{\v{s}}}, {\v{S}}tefan and {Kalman{\v{c}}ok}, Du{\v{s}}an and {Vil{\'a}gi}, Jozef and {{\v{S}}imon}, Jaroslav and {Vere{\v{s}}}, Peter and {{\v{S}}ilha}, Ji{\v{r}}{\'\i} and {Bu{\v{c}}ek}, Marek and {Gal{\'a}d}, Adri{\'a}n and {Rus{\v{n}}{\'a}k}, Patrik and {Hr{\'a}bek}, Peter and {{\v{D}}uri{\v{s}}}, Franti{\v{s}}ek and {Rudawska}, Regina},
        title = "{All-sky Meteor Orbit System AMOS and preliminary analysis of three unusual meteor showers}",
      journal = {Planetary and Space Science},
         year = 2015,
        month = dec,
       volume = {118},
        pages = {102-106},
          doi = {10.1016/j.pss.2015.07.007},
       adsurl = {https://ui.adsabs.harvard.edu/abs/2015P&SS..118..102T}
}

@ARTICLE{tsiganis05,
       author = {{Tsiganis}, K. and {Gomes}, R. and {Morbidelli}, A. and {Levison}, H.~F.},
        title = "{Origin of the orbital architecture of the giant planets of the Solar System}",
      journal = {Nature},
         year = 2005,
        month = may,
       volume = {435},
        pages = {459-461},
          doi = {10.1038/nature03539},
       adsurl = {https://ui.adsabs.harvard.edu/abs/2005Natur.435..459T}
}

@ARTICLE{vida22,
       author = {{Vida}, Denis and {Blaauw Erskine}, Rhiannon C. and {Brown}, Peter G. and {Kambulow}, Jonathon and {Campbell-Brown}, Margaret and {Mazur}, Michael J.},
        title = "{Computing optical meteor flux using global meteor network data}",
      journal = {Monthly Notices of the Royal Astronomical Society},
         year = 2022,
        month = sep,
       volume = {515},
        pages = {2322-2339},
          doi = {10.1093/mnras/stac1766},
archivePrefix = {arXiv},
       eprint = {2206.11365},
 primaryClass = {astro-ph.EP},
       adsurl = {https://ui.adsabs.harvard.edu/abs/2022MNRAS.515.2322V}
}

@ARTICLE{vida23,
       author = {{Vida}, Denis and {Brown}, Peter G. and {Devillepoix}, Hadrien A.~R. and {Wiegert}, Paul and {Moser}, Danielle E. and {Matlovi{\v{c}}}, Pavol and {Herd}, Christopher D.~K. and {Hill}, Patrick J.~A. and {Sansom}, Eleanor K. and {Towner}, Martin C. and {T{\'o}th}, Juraj and {Cooke}, William J. and {Hladiuk}, Donald W.},
        title = "{Direct measurement of decimetre-sized rocky material in the Oort cloud}",
      journal = {Nature Astronomy},
         year = 2023,
        month = mar,
       volume = {7},
        pages = {318-329},
          doi = {10.1038/s41550-022-01844-3},
archivePrefix = {arXiv},
       eprint = {2212.06812},
 primaryClass = {astro-ph.EP},
       adsurl = {https://ui.adsabs.harvard.edu/abs/2023NatAs...7..318V}
}

@ARTICLE{walsh11,
       author = {{Walsh}, Kevin J. and {Morbidelli}, Alessandro and {Raymond}, Sean N. and {O'Brien}, David P. and {Mandell}, Avi M.},
        title = "{A low mass for Mars from Jupiter's early gas-driven migration}",
      journal = {Nature},
         year = 2011,
        month = jul,
       volume = {475},
        pages = {206-209},
          doi = {10.1038/nature10201},
archivePrefix = {arXiv},
       eprint = {1201.5177},
 primaryClass = {astro-ph.EP},
       adsurl = {https://ui.adsabs.harvard.edu/abs/2011Natur.475..206W}
}

\end{IMCpaper}
\end{document}